\documentclass[aps,prb,twocolumn,superscriptaddress,showpacs]{revtex4}
\usepackage{amsmath}
\usepackage{amssymb}
\usepackage{graphicx}
\usepackage{dcolumn}
\usepackage{bm}

\begin{document}

\title{Conducting Jahn-Teller Domain Walls in Undoped Manganites. }
\author{ Juan Salafranca}
\affiliation{Department of Physics and Astronomy, University of
  Tennessee, Knoxville, Tennessee 37996, USA}
\affiliation{Materials Science and Technology Division, Oak Ridge
  National Laboratory, Oak Ridge, Tennessee 32831, USA}
\author{Rong Yu}
\affiliation{Department of Physics and Astronomy, Rice University, Houston, Texas 77005, USA}
\author{Elbio Dagotto}
\affiliation{Department of Physics and Astronomy, University of
  Tennessee, Knoxville, Tennessee 37996, USA} 
\affiliation{Materials Science and Technology Division, Oak Ridge
  National Laboratory, Oak Ridge, Tennessee 32831, USA} 

\email{jsalafra@utk.edu}


\begin{abstract}  
We investigate the electronic properties of multi-domain
configurations 
in models for undoped manganites by means of variational and Monte Carlo
techniques. These materials display simultaneous Jahn-Teller
distortions and magnetic ordering. We find that a band of
electronic states appears associated with  Jahn-Teller domain
walls, and this band is localized in the direction perpendicular to 
the walls. The energy and width of this band depends on the 
conformational properties of the domain walls. At finite temperatures, 
the conductance along the domain walls, induced by the localized 
domain wall bands,  is orders magnitude larger than in the bulk. 
\end{abstract}

\date{\today}

\maketitle
\section{introduction}

In the quest to understand and control the properties of broken
symmetry phases,  
domain walls play a central role. Among the many materials with ground
state broken symmetries, strongly correlated transition metal
oxides, and manganites among them,  are 
particularly interesting. In these compounds several degrees of freedom are
simultaneously important, and different broken symmetry phases with similar
characteristic energy scales either compete, as in phase 
separated materials,\cite{Dagotto_Science}  or coexist, as it occurs in
multiferroics.~\cite{Mostovoy_multiferroics,Scott_multiferroics} 
Research on magnetic domain walls in general has been particularly intense,
and it has proven to be extremely important to explain the static and dynamic properties of
magnetic materials.  Early
work~\cite{KittelRMP,LandauEDM} led to the development of a variety of  important 
concepts to understand domain walls. 
Technological applications, and the need to control and understand the
important details of domain walls, stimulated considerable and wide  
research on this topic,\cite{DW1,DW2,DW3,Marrows_domainwalls} unveiling
a variety of interesting aspects of these walls.  
In particular, the magnetic domain walls of manganites, the materials with the
colossal magnetoresistance, have also attracted much
attention. Already in Ref.~\onlinecite{Schiffer}, the large
magnetoresistance was attributed to domain wall
scattering, an hypothesis that led to both experimental and
theoretical work on the subject
of resistance of domain walls. 
\cite{Li_DWR,Levy_DWR,Mathur_DWR,Suzuki_DWR,Golosov2003DEDW,Nagaosa_DWR,STM_DWR,FLORIDA_DWR}
Recently, it was shown that magnetic domain walls in a ferromagnetic
metallic material could be
insulating.\cite{insulatingDWs}

Similarly, domain walls
and gradients of the order parameter play a crucial role in
our understanding of other collective phenomena, such as
superconductivity,~\cite{GinzburgLandau} and 
ferroelectrity.\cite{domainsPTO,Junquera_closure} 
Early theoretical work showed that physical properties absent in
 bulk materials can arise in domain walls,\cite{Privratska} and the conductivity of
 ferroelastic domain walls has been showed to be different from the
 conductivity in the bulk.\cite{Arid,Bartels} 
Another very interesting field is the
study of properties of domain walls in multiferroics, or magnetoelectric
materials. Different orderings can change across domain walls in these
materials, and indeed it has been observed than often the change in one
order parameter is correlated with modifications in another.\cite{Fiebig2002clampedDWs}
The interplay between order parameters can affect the physical
properties of domain walls. In the antiferromagnetic and ferroelectric phases of hexagonal
HoMnO$_3$,\cite{Lottermoser2004DWsHoMnO3} there is a 
transition between two symmetry nonequivalent antiferromagnetic 
phases. Wide domain walls where the spins rotate smoothly
appear close to this transition,  
producing a pronounced magnetoelectric
behavior. Interestingly, this behavior is not observed associated with
the abrupt domain walls where magnetization reversal
is coupled to a change in the ferroelectric order parameter.  
In helicoidal magnetic manganites, with a perovskite crystal
structure, the study of the dielectric dispersion and its behavior varying 
the temperature\cite{Tokura2002dynamicsDW} allowed to identify the motion of
domain walls as the origin of the substantial enhancement of the dielectric
constant in materials like DyMnO$_3$ and TbMnO$_3$. 

\begin{figure}
  \includegraphics[width=0.9\columnwidth,clip]{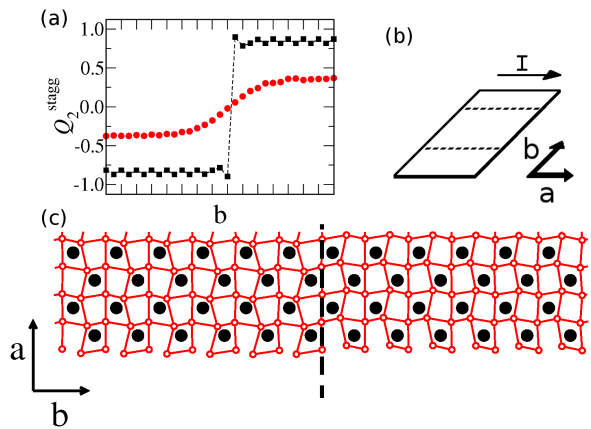}
\caption{Jahn-Teller domain wall at $T$=0, as obtained by the classical
  variables minimization process described in the text. The cluster
  size used is 4$\times$28. (a) Variation of the order parameter
  perpendicular to the domain wall, (black squares are results for $\lambda$=1.25; red
  circles are for $\lambda$=1). 
  (b) Geometry for the calculation of the conductance, see text.
  (c) 2$\times$12 portion of the simulation cluster near a domain wall
  showing octahedra distortions in real space, for $\lambda$=1.25. The
  oxygen  displacements obtained by energy minimization have been
  rescaled to clearly show the two different lattice configurations 
left and right of the domain wall. } 
\label{Fig:SDW}
\end{figure}

Of particular motivational interest for the work described here is Ref.~\onlinecite{Seidel},
where manipulation and characterization of domain walls in 
ferroelectric and antiferromagnetic BiFeO$_3$ was demonstrated.
The authors of that effort could create ferroelectric domain
walls and showed that domain walls with particular orientations exhibited larger
conductance than others, within an otherwise good insulating
material. Contrary to previous work, the authors focused on electronic
transport in the direction parallel to the domain walls, instead of
transport across them as in other efforts 
in manganites.\cite{Li_DWR,Levy_DWR,Mathur_DWR,Suzuki_DWR,Golosov2003DEDW,Nagaosa_DWR,STM_DWR,FLORIDA_DWR} 
A very recent work studied domain walls in a non-perovskite oxide.\cite{Choi}
Domain walls in hexagonal YMnO$_3$  have been shown to be more
insulating than in the bulk, an effect actually opposite to the
BiFeO$_3$ case.
Here, motivated by those previous efforts, we study the electronic properties of domain walls
in undoped perovskite manganites. These materials are not only
antiferromagnetic insulators, but also have a phase transition above
room temperature, where an ordered pattern of distorted MnO$_6$ octahedra    
sets in. We show that at finite temperature ($T$),
structural domain walls  display a conductance orders of
magnitude larger that the conductance observed in bulk materials.
  
In particular, we concentrate on the insulating, Jahn Teller (JT) ordered phase
of undoped manganites with an A-type spin antiferromagnetic
order.\cite{rodriguezcarvajalLaMnO3,hotta_2003a}
Our standard variational and Monte Carlo calculations, using well-tested and reliable
models, predict that: {\it (i)} {\it domain wall electronic states} exist,
and are localized in the direction perpendicular to the domain
wall. They are  
analogous to surface states and appear associated with the presence of an
structural domain wall. {\it (ii)} These states form narrow bands within the
energy gap of the bulk material. The periodicity of the structural
distortions determine the position and width of theses bands. {\it (iii)} At
low temperature $T$ there is a  gap between the domain wall bands, and as
$T$ increases, conformational changes in the domain wall make the
localized bands wider, and induce some spectral weight at the Fermi
energy, leading to a finite conductance in the
direction parallel to the domain walls.

\section{Model and Techniques}

\subsection{Model Hamiltonian}
\label{subsecH}

The Hamiltonian used here is given by:

\begin{equation}
H=H_{\rm DE}+H_{\rm e-ph}
\label{H}
\end{equation}

$H_{\rm  DE}$ is the well-known two-orbital double exchange model Hamiltonian. This model
describes the kinetic energy of the $e_{\rm g}$ electrons and their
interaction with the magnetic background of the $t_{\rm 2g}$ core
spins.\cite{Dagotto_book} More precisely, it is given by
\begin{equation}
H_{\rm DE}=- \sum_{i,j,\gamma,\gamma'} f_{i,j} t^u_{\gamma,\gamma'} c_{i,\gamma}^{\dagger} c_{j,\gamma'}.
\label{HTB}
\end{equation}
Here $c_{i,\gamma}^{\dagger}$ creates an electron at the Mn site $i$,
in the $e_{\rm g}$ orbital $\gamma$ ($\gamma=1,2$ with $1=|x^2-y^2 \rangle$
and $2=|3 z^2-r^2 \rangle$). The factor $f_{i,j}$ affecting the hopping amplitudes in the limit
of an infinite Hund's coupling depends
on the Mn core spins orientation given by the angles $\theta$ and
$\psi$ via   the double-exchange mechanism {\small
$f_{i,j}=\cos(\theta_i/2) \cos(\theta_j/2)+
\exp[i(\psi_i-\psi_j)]\sin(\theta_i/2) \sin(\theta_j/2)$}. 
The actual hopping amplitudes $t$ take into account the different overlaps 
between the orbitals along the directions u=$x$,$y$:~\cite{SlaterKoster} $t _{
  1,1}^x$=$3t _{2,2}^x$=$\sqrt{3}t _{1,2}^x$=$t$ and
 $t _{1,1}^y$=$3t _{2,2}^y$=-$\sqrt{3}t _{1,2}^y$=$t$. 
$t$ is taken as the energy unit throughout this work, and its value depends on
material details. Comparison between theoretical calculations and
experiments suggests that its value is of the order of half an electron volt.\cite{Fermiarcs}

The phononic portion of the
Hamiltonian reads:
\begin{eqnarray}
H_{\rm e-ph}=
 \lambda \sum _{i} \left ( - Q _{1i} \rho _i + Q _{2i} \tau _{xi}
+ Q _{3i} \tau _{zi} \right )\nonumber \\
+ \frac{1}{2} \sum _i \left (  \beta Q _{1i} ^2+ Q
^2_{2i} + Q _{3i}^2 \right ) \, \, \, ,
\label{HJT}
\end{eqnarray}
where $\rho_i$ is the density operator at site $i$, and $\tau _{xi} $ and $  \tau
_{zi}$ are the corresponding Pauli matrices in the {\it e}$_{\rm g}$ subspace, that
express the coupling of the lattice distortions, $Q$, to the electrons.
In particular, $Q_{1i}$ is the breathing mode of the MnO$_6$ octahedra
around the $i$-th 
manganese ion, and $Q_{2i}$, $Q_{3i}$ are the Jahn Teller modes.
\cite{Dagotto_book} $\beta$ = 2 is the stiffness of the breathing mode
that effectively takes into account suppressed charge fluctuations due to
electron-electron interactions.\cite{Hotta_hartree}

\subsection{Calculation Method}

Undoped manganites, such as LaMnO$_3$, crystallize in an orthorhombic
structure, with two nonequivalent manganese positions. Hamiltonian
(\ref{H}) has cubic symmetry, but in order to better compare with
experiments we maintain the axis and notation corresponding to the
orthorhombic unit cell, in particular we call $b$ axis the direction
perpendicular to the domain walls (see Fig~\ref{Fig:SDW}). We focus on
the antiferromagnetic $A$-type phase,\cite{Wollan}
experimentally found for $La$ manganite and other trivalent
cations of similar size.\cite{rodriguezcarvajalLaMnO3} 
In this  A-phase there are ferromagnetic planes
coupled antiferromagnetically in 
the $c$-axis direction. Double exchange effectively decouples electrons in the $ab$
atomic planes, and 2D systems provide a good description of the physics of these
materials. In real compounds, certainly a coupling exists along the $c$ axis
that prevents 2D fluctuations to dominate the magnetic and electronic
properties of these compounds.

The Hamiltonian (\ref{H}) depends on the octahedra and core spin
configurations, ${Q_i}$ and ${\vec S_i}$. These have been
approximated as classical degrees of freedom, as in most of the
theoretical literature on manganites,  and they determine the
electronic properties of the system. At finite $T$, Monte Carlo
simulations allow us to obtain the relevant equilibrium configurations and
calculate thermal averages. ${Q_i}$ and ${\vec S_i}$ are used as the
Monte Carlo variables and the resulting quadratic Hamiltonian for each
configuration is exactly diagonalized using standard numerical
routines.\cite{lapack} Further details about the Monte Carlo method
and its application to manganites can be found 
in Refs. \onlinecite{Dagotto_book} and \onlinecite{SalafrancaMonte}. 

At low enough $T$, only one classical configuration becomes relevant, and it can be
determined by a suitable minimization algorithm. The results for $T$=0
presented here have been obtained with the Broyden method.
\cite{Recipes} Although it 
still involves the diagonalizing of the fermionic sector for each
step, this minimization process is faster than the standard 
Monte Carlo method and it can be
applied to calculate results on larger clusters. Comparison between results
obtained with the two methods shows that size effects are small. 
This calculation method has been successfully applied to study
uniform phases in manganites.\cite{Sen_04,hotta_2003a,Fermiarcs}
Here, with the appropriate boundary 
conditions discussed below, we study the properties of structural domain
walls in undoped Manganites.

\begin{figure}
\includegraphics[width=0.95\columnwidth,clip]{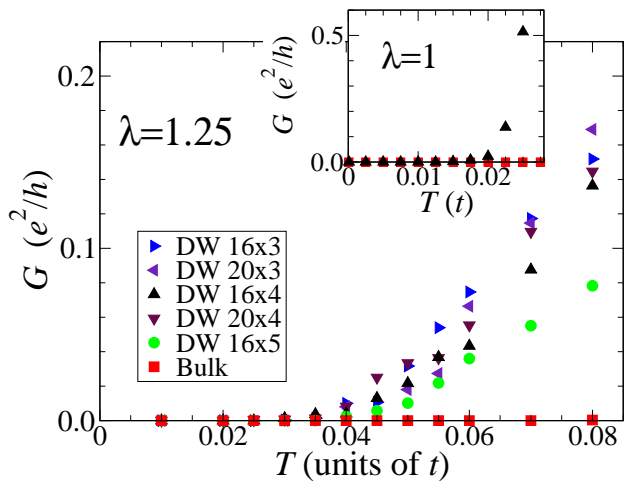}
\caption{Conductance as a function of temperature for different
  system sizes. {\it DW} stands for conductance along domain walls (see
  Fig.~\ref{Fig:SDW}) and {\it Bulk}, for conductance in the bulk limit. 
Details about the Monte Carlo procedure are
  provided in the text. Despite some small size effects, the conductance is
  clearly qualitative different when domain walls are present. For the
  bulk case, data for all sizes fall 
  within the same curve in the scale of the figure. 
  Note that the conductance is roughly independent of the
  dimensions along the $b$ axis (perpendicular to the walls), consistent with the 
hypothesis that the 
  conductance takes place along the domain walls. Also note that assuming a
hopping amplitude $t$ of approximately 0.5 eV, then $T=0.06t$ is approximately room $T$.} 
\label{Fig:GvsT}
\end{figure}

The ground state configuration for Hamiltonian (\ref{H}) with one electron per site
is a spin ferromagnetic plane with an ordered pattern of $Q_2$ distortions characterized
by a ($\pi$, $\pi$) wave vector in the 2D cubic notation.
\cite{hotta_2003a,incommensurate} These 
orderings  correspond in 3D to the experimentally observed A-type antiferromagnetic phase,
and the ($\pi$, $\pi$, 0) JT ordered phase.\cite{rodriguezcarvajalLaMnO3} In the
orthorhombic notation, the two octahedra around the two Mn ions in the unit
cell have opposite $Q_2$ distortions, one is elongated along the $x$ axis, and
the other along the $y$ axis. The two possible ground state configurations 
intrinsic of a staggered order parameter (one reachable from the other by
merely a global shift in one lattice spacing in any axes direction) are
obtained by repeating the two possible units in the $a$ and $b$ directions. 
We label $Q_2^{stagg}$ the order parameter, whose value is the
magnitude of the distortions and its sign distinguishes between the
two possible configurations.

Monte Carlo or optimization calculations at low temperatures with
standard periodic boundary conditions converge to one of the two
possible ground state configurations. In order to study the 
domain walls that we wish to focus on in this effort   
different boundary conditions have to be imposed. In a $M$$\times$$N$ cluster,
we fixed sites with coordinate $b$=1 and $b$=$N$ to the bulk equilibrium values
corresponding to  $Q_2^{stagg}>$0, while sites at the center layer $b$=$N/2$ are
constrained to have distortions with the same magnitude but
corresponding to the other configuration $Q_2^{stagg}<$0. This way, two
JT domain walls do appear in the system, and periodic boundary
conditions can still be used for the fermionic sector.

The high computational cost of the repeating diagonalizing step of the Monte Carlo
algorithm limits the sizes of the simulation cells.  The largest cell
used in this work is $a$=4, $b$=20, which 
contains 160 Mn ions (there are two Mns in the orthorhombic unit cell).
Thermalization is achieved by 2,000 Monte Carlo steps
(as defined in Ref.~\onlinecite{SalafrancaMonte}) and measurement 
averages of 3,000 Monte Carlo configurations are used, taken every three
steps to reduce self-correlations. This computational effort 
corresponds to roughly two days of running in a typical workstation
for a 4$\times$16 system and five days for a 4$\times$20 system, 
for each fixed set of parameters in the problem.

The conductance is calculated within the Landauer
formalism, as explained in Ref.~\onlinecite{VergesCPC}. To
minimize size effects, we calculate the Green function of the system
attached to leads that are identical copies of itself. Leads are
enlarged by adding more copies of the system until the Green function
converges, providing an 
efficient method to calculate the intrinsic
conductance.\cite{SalafrancaMonte} As illustrated in
Fig.~\ref{Fig:SDW} the conductance of domain walls refers to conductance
evaluated in the direction parallel to the domain walls.  

\section{results}

\begin{figure}
\includegraphics[width=0.9\columnwidth,clip]{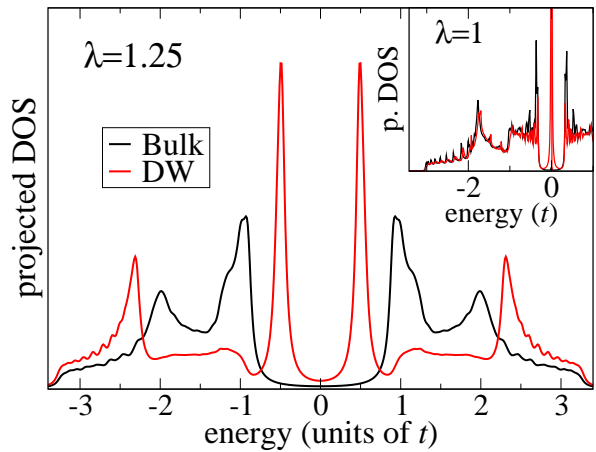} 
\caption{Density of states a $T$=0. States appear inside the bulk gap
  energy region when close to a domain wall (more specifically DW refers to
the projected DOS at the line of manganese sites the closest to the domain wall
in Fig.~1(c)). 
The states in this band are
  localized in the 
  direction perpendicular to the domain wall (their weight decays exponentially
  in that direction), but a small dispersion
  exists parallel to the
  domain wall. The band arising from the two $e_{\rm g}$ orbitals
  is split due to the periodicity of the JT distortions in the
  direction parallel to the domain wall.  
  For $\lambda$=1.25 the gap is of the order of $t$ ($\approx$0.5 eV), even in the
  presence of a domain wall. For the smoother domain wall
  corresponding to $\lambda$=1 the gap is very small and it cannot be resolved in this
  scale (inset).} 
\label{Fig:DOST0}
\end{figure}
Before addressing the results regarding electronic transport, let us
discuss the nature  of the Jahn Teller domain walls. Figure
\ref{Fig:SDW}(a)  presents the variation of the order parameter as a
function of distance  
across the domain wall. In magnetic domain walls, it is well known\cite{KittelRMP}
that the width varies as $\sqrt{D/A}$, where $A$ is
the magnetic anisotropy and $D$ is the spin-stiffness.
In the case of an isolated Jahn Teller center, 
the energy depends on ($Q_2^2$+$Q_3^2$) and the
anisotropy is zero. The cooperative nature of Jahn Teller effect in
manganites and the competition with the kinetic energy double-exchange 
term (even in the
insulating state) gives rise to an effective anisotropy that depends on the value of
$\lambda$. These same effects stabilize the $Q_2$ order in the
$ab$ plane.\cite{Satpathy}   Therefore, $\lambda$/$t$ controls the
thickness of the domain 
wall. Figure~\ref{Fig:SDW} (a) shows how an abrupt domain wall is the
minimum energy configuration for $\lambda$=1.25, but for $\lambda$=1,
the domain wall expands over several unit cells. The first $\lambda$ value provides a
more accurate description of the real materials, as it can be deduced
by the magnitude of the electronic gap discussed later.  A real-space
view of the octahedra distortions in the $ab$ plane for the
$\lambda$=1.25 case is presented in Fig. \ref{Fig:SDW}(c).

The conductance vs. $T$ plots contained in Fig.~\ref{Fig:GvsT} constitute the main results of this
manuscript. The conductance, $G$, refers to the conductance in the
direction parallel to 
the domain wall, as shown in Fig.~\ref{Fig:SDW}(b), and it is provided as
a function of $T$.  These
results have been obtained by means of Monte Carlo simulations with
different system sizes, as indicated in
Fig.~\ref{Fig:GvsT}. 
$G$ is essentially zero (within our numerical precision) for bulk 
configurations, when no domain walls are present, for all the $T$s
considered in Fig.~\ref{Fig:GvsT}. However, when domain walls are
present in the 
system, the conductance shows a clear upturn when increasing the temperature. 
For $\lambda$=1, where the
domain wall is fairly wide,  this upturn takes 
place at $T\approx$0.02, while a larger $T \approx$0.05 is needed to induce a
significant value for $G$ in the  $\lambda$=1.25 case. There are some size
effects in the results, but the increase with temperature of the
conductance along domain 
walls, as compared to its negligible bulk value, is clearly present for all system
sizes within our computational capabilities.

Figure~\ref{Fig:DOST0} shows the density of states (DOS) projected over
different sites at zero $T$. It has been obtained by providing to the 
eigenenergies arising from the exact diagonalization procedure a small
lorentzian width (0.01 $t$). 
For the bulk case (no domain walls) the DOS
has a large gap due to Jahn Teller effects.  This gap
scales with $\lambda$: it is roughly 2$t$ for $\lambda$=1.25 and half
that value for  $\lambda$=1. Considering that $t$ is in the range
0.5-1 eV, $\lambda$=1.25 is more consistent with the experimental
estimations of the gap (1.6 eV in Ref.~\onlinecite{gapLMO1} and  2.5
eV in Ref.~\onlinecite{gapLMO2}).

\begin{figure}
\includegraphics[width=0.9\columnwidth,clip]{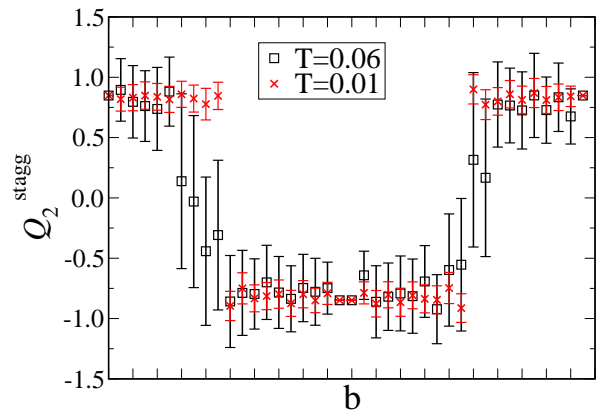}
\caption{JT domain wall profile for $T$=0.01 and $T$=0.06 obtained via
  Monte Carlo simulations, at $\lambda$=1.25 and for an $a$=3 $b$=20 system. 
The error bars show standard
  deviation due to thermal fluctuations. The temperature makes the domain
  walls smoother, reducing the energy gap. }  
\label{Fig:DWfT}
\end{figure}

When the density of states is projected over sites further than $\approx$
4 unit cells from the domain wall, we recover essentially the bulk density
of states. However, a projection over Mn sites close to the domain wall is more
interesting. Figure~\ref{Fig:DOST0} shows that the DOS the closest to the 
domain
wall has some features that do not appear in the bulk region. Two
narrow peaks appear at energies located within the bulk energy
gap. These two peaks are separated in energy creating a new, smaller,
gap that depends strongly on $\lambda$. It is $\approx t/10$ for 
$\lambda$=1 and increases to $\approx t$ for $\lambda$=1.25.  The integrated
weight of these peaks decays exponentially in the $b$ direction,
perpendicular to the domain wall, so these states are localized at the domain walls.
A least-squares fitting shows that
the decay length is of the order of the nearest neighbors
distance. The fermionic wave functions coming out of the simulations, via the
exact diagonalization procedure, have a well
defined momentum in the $a$ direction.    
Therefore, we can state that the new peaks correspond to a band localized
in the $b$ direction and 
with a small dispersion ($\approx t/20$) parallel to the domain wall. Note that
the artificial broadening 0.01~$t$ is not sufficient to understand the width
of the in-gap peaks; in fact,  
the actual dispersion in the $a$ direction produces their
intrinsic width.

These bands are obvious candidates to explain the conductance
results in Fig.~\ref{Fig:GvsT}. Although the energy gap due to the
splitting of the localized band is rather small for $\lambda$=1, it is
too large to explain the nonzero conductance at $T$=0.05
observed for the more realistic 
case $\lambda$=1.25. As the electronic
states associated with the domain wall disperse in 
the $a$ direction, some splitting due to the periodic Jahn Teller
distortions in that direction can be expected. This explains why the
splitting decreases with $\lambda$. Note that the conductance
calculation has enough precision to determine that the small gap in
the $\lambda$=1 case, difficult to appreciate in Fig.~\ref{Fig:DOST0},
still suppresses transport at the lowest temperatures.

In order to understand the onset of electronic transport with
temperature, we examine the structure of the domain walls as $T$ 
increases (Fig.~\ref{Fig:DWfT}), for $\lambda$=1.25. We have observed
that the order parameter changes much more 
abruptly at $T$=0.01 as compared to $T$=0.06. The points in
Fig.~\ref{Fig:DWfT} correspond to average values over 1,500 Monte Carlo
steps, with the error bars showing the standard effect of thermal
fluctuations. The domain wall is clearly wider for $T$=0.06, with a width that
effectively corresponds to a smaller $\lambda$ if the calculation were done at very low $T$ (as discussed below). Entropy reduces the effective
anisotropy (or lowers the order parameter stiffness). A visual inspection of
Monte Carlo snapshots, namely equilibrated configurations of classical variables, 
shows that the widening of 
the domain wall is due to changes in the intrinsic width, and the
contribution from thermal
vibrations of the center of the wall is small.

\begin{figure}
\includegraphics[width=0.9\columnwidth,clip]{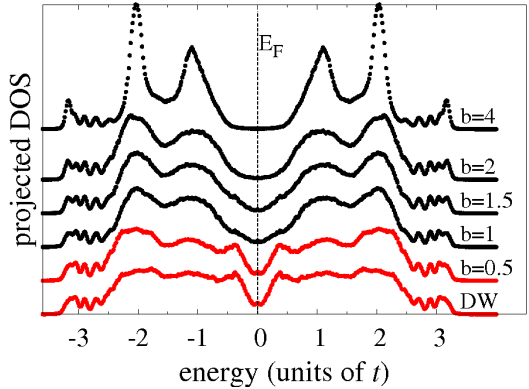}
\caption{Thermal averaged electronic density of states for $T$=0.06, $\lambda=1.25$, projected over
  different atomic planes parallel to the domain wall (in practice these are atomic
chains since we use a 2D geometry). The zero of reference for each case is provided
by the result at the far left or right in the figure. 
At finite temperature
  the domain walls are not as abrupt as at $T$=0, the bands associated with it (see
  Fig.~\ref{Fig:DOST0}) are wider and some spectral weight is shifted
  to the Fermi energy $E_{\rm F}$. The eigenenergies arising from the Monet Carlo
  simulation have been given a small width (0.01$t$), and the
  spectrum has been symmetrized with respect  
  to $E_{\rm F}$ to reduce noise. 
} 
\label{Fig:PDOS_fT}
\end{figure}

The widening of the domain wall induces changes in the density of
states. In Fig.~\ref{Fig:PDOS_fT} we show the DOS projected on
sites at different distances from the domain wall, thermally
averaged at $T$=0.06. Particle-hole symmetry was assumed, and the
spectrum symmetrized with respect to $E_{\rm F}=0$ in order to
reduce the noise (Monte Carlo calculations are more noisy than optimization results due to
thermal and finite size effects). However, it can be clearly
observed how the band associated with the domain wall is now wider
in energy, and has some spectral weight at the Fermi energy, that is
responsible for the finite conductance. Notice that this band is still
localized in the $b$ direction and by moving 4 sites away from the
domain wall, the bulk density of states is recovered.

An intuitive way to understand the results presented above is to
consider that the coupling $\lambda$ is now effectively smaller near
the domain wall. But note that temperature must also be incorporated. At $T$=0,
the domain wall is abrupt for the realistic coupling $\lambda$=1.25,
and therefore this coupling produces distortions at the domain wall
as large as in the rest of the system. The bands appear
associated with the domain wall due to the change in periodicity. 
For smaller $\lambda$'s, domain walls are smoother. Wider domain
walls have a smaller gap, relative to the bulk gap. For
$\lambda$=1.25, the gap between the domain walls gaps is half the bulk
value, while is an order of magnitude smaller than the bulk gap for
$\lambda$=1. Thus, $T$ must also be included to justify the 
effective reduction of the value of the electron lattice 
coupling. From the
structural point of view, this effect results in a smaller anisotropy and an
increase in the domain wall width. As regarding the electronic
structure, it reduces the gap between the localized bands, eventually
closing it.

\section{Conclusions} 
     
    Stimulated by the recent experimental characterization and
    manipulation of structural domain walls in a perovskite transition metal
    oxide,\cite{Seidel} we have undertaken the
    theoretical study of domain walls in undoped manganites. These
    materials display an ordered phase with an alternating pattern of
    distorted MnO$_6$ octahedra, and mobile $e_{\rm g}$ electrons are known to couple
    strongly to these
    distortions. The equilibrium configurations and electronic
    properties of structural domain walls have been examined, 
    and they remarkably differ form the properties of the bulk.
    Since it is an important characteristic for applications and
    a property directly measurable by experiments, here special attention
    has been paid to the electronic conductance of the system.

    Electronic transport along structural domain walls is dominated by 
    two narrow electronic bands associated with them. These domain walls
    electronic  states have energies within the bulk energy gap
    region. At low $T$'s, domain walls are abrupt, and the narrow bands
    are separated in energy. The energy gap between them,
    although smaller than the bulk energy gap, is still large 
enough to make the conductance
    negligible at low $T$. However, for $T$'s of the order of
    room temperature, thermal fluctuations favor wider domain walls. This makes
    the dispersion of the domain walls electronic states larger, reducing 
the energy gap, and inducing some spectral weight at the Fermi
    energy. Correspondingly, conductance along domain walls is several
    orders of magnitude larger for the temperature range studied than in the bulk, while it is
    still zero within our numerical precision for the bulk phase. 
    Our results pose the question of whether  the
    conductance enhancement by domain walls electronic states induced by changes in the
    periodicity of the system might be a more general
    phenomenon, and suggest the possibility that they might play a role
    in recent experiments showing conducting domain walls in a multiferroic
    perovskite oxide.\cite{Seidel} These issues certainly 
deserve further theoretical and experimental studies.

\acknowledgments
The authors acknowledge useful conversations with Jan Seidel. 
This work was supported by the NSF grant DMR-0706020 and the
Division of Materials Science and Engineering, Office of Basice Energy
Sciences, 
U.S. Department of Energy. 
R.Y. acknowledges support from NSF Grant No. DMR-0706625, the Robert
A. Welch Foundation Grant No. C-1411, and the W.
M. Keck Foundation. 

\bibliography{bibliography}

\begin{thebibliography}{46}
\expandafter\ifx\csname natexlab\endcsname\relax\def\natexlab#1{#1}\fi
\expandafter\ifx\csname bibnamefont\endcsname\relax
  \def\bibnamefont#1{#1}\fi
\expandafter\ifx\csname bibfnamefont\endcsname\relax
  \def\bibfnamefont#1{#1}\fi
\expandafter\ifx\csname citenamefont\endcsname\relax
  \def\citenamefont#1{#1}\fi
\expandafter\ifx\csname url\endcsname\relax
  \def\url#1{\texttt{#1}}\fi
\expandafter\ifx\csname urlprefix\endcsname\relax\def\urlprefix{URL }\fi
\providecommand{\bibinfo}[2]{#2}
\providecommand{\eprint}[2][]{\url{#2}}

\bibitem[{\citenamefont{Dagotto}(2005)}]{Dagotto_Science}
\bibinfo{author}{\bibfnamefont{E.}~\bibnamefont{Dagotto}},
  \bibinfo{journal}{Science} \textbf{\bibinfo{volume}{309}},
  \bibinfo{pages}{257} (\bibinfo{year}{2005}).

\bibitem[{\citenamefont{Cheong and Mostovoy}(2007)}]{Mostovoy_multiferroics}
\bibinfo{author}{\bibfnamefont{S.}~\bibnamefont{Cheong}} \bibnamefont{and}
  \bibinfo{author}{\bibfnamefont{M.}~\bibnamefont{Mostovoy}},
  \bibinfo{journal}{Nature Materials} \textbf{\bibinfo{volume}{6}},
  \bibinfo{pages}{13} (\bibinfo{year}{2007}).

\bibitem[{\citenamefont{Eerenstein et~al.}(2006)\citenamefont{Eerenstein,
  Mathur, and Scott}}]{Scott_multiferroics}
\bibinfo{author}{\bibfnamefont{W.}~\bibnamefont{Eerenstein}},
  \bibinfo{author}{\bibfnamefont{N.}~\bibnamefont{Mathur}}, \bibnamefont{and}
  \bibinfo{author}{\bibfnamefont{J.}~\bibnamefont{Scott}},
  \bibinfo{journal}{Nature} \textbf{\bibinfo{volume}{442}},
  \bibinfo{pages}{759} (\bibinfo{year}{2006}).

\bibitem[{\citenamefont{Kittel}(1949)}]{KittelRMP}
\bibinfo{author}{\bibfnamefont{C.}~\bibnamefont{Kittel}},
  \bibinfo{journal}{Rev. Mod. Phys.} \textbf{\bibinfo{volume}{21}},
  \bibinfo{pages}{541} (\bibinfo{year}{1949}).

\bibitem[{\citenamefont{Lifshitz et~al.}(1984)\citenamefont{Lifshitz, Landau,
  and Pitaevskii}}]{LandauEDM}
\bibinfo{author}{\bibfnamefont{E.}~\bibnamefont{Lifshitz}},
  \bibinfo{author}{\bibfnamefont{L.}~\bibnamefont{Landau}}, \bibnamefont{and}
  \bibinfo{author}{\bibfnamefont{L.}~\bibnamefont{Pitaevskii}},
  \emph{\bibinfo{title}{{Electrodynamics of continuous media}}}
  (\bibinfo{publisher}{Pergamon Press}, \bibinfo{year}{1984}).

\bibitem[{\citenamefont{Marrows}({2005})}]{Marrows_domainwalls}
\bibinfo{author}{\bibfnamefont{C.}~\bibnamefont{Marrows}},
  \bibinfo{journal}{Adv.\ Phys.} \textbf{\bibinfo{volume}{{54}}},
  \bibinfo{pages}{{585}} (\bibinfo{year}{{2005}}).

\bibitem[{\citenamefont{Tatara and Kohno}(2004)}]{DW1}
\bibinfo{author}{\bibfnamefont{G.}~\bibnamefont{Tatara}} \bibnamefont{and}
  \bibinfo{author}{\bibfnamefont{H.}~\bibnamefont{Kohno}},
  \bibinfo{journal}{Phys. Rev. Lett.} \textbf{\bibinfo{volume}{92}},
  \bibinfo{pages}{086601} (\bibinfo{year}{2004}).

\bibitem[{\citenamefont{Kl\"aui et~al.}(2005)\citenamefont{Kl\"aui, Jubert,
  Allenspach, Bischof, Bland, Faini, R\"udiger, Vaz, Vila, and Vouille}}]{DW2}
\bibinfo{author}{\bibfnamefont{M.}~\bibnamefont{Kl\"aui}},
  \bibinfo{author}{\bibfnamefont{P.-O.} \bibnamefont{Jubert}},
  \bibinfo{author}{\bibfnamefont{R.}~\bibnamefont{Allenspach}},
  \bibinfo{author}{\bibfnamefont{A.}~\bibnamefont{Bischof}},
  \bibinfo{author}{\bibfnamefont{J.~A.~C.} \bibnamefont{Bland}},
  \bibinfo{author}{\bibfnamefont{G.}~\bibnamefont{Faini}},
  \bibinfo{author}{\bibfnamefont{U.}~\bibnamefont{R\"udiger}},
  \bibinfo{author}{\bibfnamefont{C.~A.~F.} \bibnamefont{Vaz}},
  \bibinfo{author}{\bibfnamefont{L.}~\bibnamefont{Vila}}, \bibnamefont{and}
  \bibinfo{author}{\bibfnamefont{C.}~\bibnamefont{Vouille}},
  \bibinfo{journal}{Phys. Rev. Lett.} \textbf{\bibinfo{volume}{95}},
  \bibinfo{pages}{026601} (\bibinfo{year}{2005}).

\bibitem[{\citenamefont{Tatara and Fukuyama}(1997)}]{DW3}
\bibinfo{author}{\bibfnamefont{G.}~\bibnamefont{Tatara}} \bibnamefont{and}
  \bibinfo{author}{\bibfnamefont{H.}~\bibnamefont{Fukuyama}},
  \bibinfo{journal}{Phys. Rev. Lett.} \textbf{\bibinfo{volume}{78}},
  \bibinfo{pages}{3773} (\bibinfo{year}{1997}).

\bibitem[{\citenamefont{Schiffer et~al.}(1995)\citenamefont{Schiffer, Ramirez,
  Bao, and Cheong}}]{Schiffer}
\bibinfo{author}{\bibfnamefont{P.}~\bibnamefont{Schiffer}},
  \bibinfo{author}{\bibfnamefont{A.~P.} \bibnamefont{Ramirez}},
  \bibinfo{author}{\bibfnamefont{W.}~\bibnamefont{Bao}}, \bibnamefont{and}
  \bibinfo{author}{\bibfnamefont{S.-W.} \bibnamefont{Cheong}},
  \bibinfo{journal}{Phys. Rev. Lett.} \textbf{\bibinfo{volume}{75}},
  \bibinfo{pages}{3336} (\bibinfo{year}{1995}).

\bibitem[{\citenamefont{Golosov}(2003)}]{Golosov2003DEDW}
\bibinfo{author}{\bibfnamefont{D.~I.} \bibnamefont{Golosov}},
  \bibinfo{journal}{Phys. Rev. B} \textbf{\bibinfo{volume}{67}},
  \bibinfo{pages}{064404} (\bibinfo{year}{2003}).

\bibitem[{\citenamefont{Li et~al.}(2001)\citenamefont{Li, Hu, and
  Wang}}]{Li_DWR}
\bibinfo{author}{\bibfnamefont{Q.}~\bibnamefont{Li}},
  \bibinfo{author}{\bibfnamefont{Y.}~\bibnamefont{Hu}}, \bibnamefont{and}
  \bibinfo{author}{\bibfnamefont{H.}~\bibnamefont{Wang}},
  \bibinfo{journal}{Journal of Applied Physics} \textbf{\bibinfo{volume}{89}},
  \bibinfo{pages}{6952} (\bibinfo{year}{2001}).

\bibitem[{\citenamefont{Levy and Zhang}(1997)}]{Levy_DWR}
\bibinfo{author}{\bibfnamefont{P.~M.} \bibnamefont{Levy}} \bibnamefont{and}
  \bibinfo{author}{\bibfnamefont{S.}~\bibnamefont{Zhang}},
  \bibinfo{journal}{Phys. Rev. Lett.} \textbf{\bibinfo{volume}{79}},
  \bibinfo{pages}{5110} (\bibinfo{year}{1997}).

\bibitem[{\citenamefont{Mathur et~al.}(1999)\citenamefont{Mathur, Littlewood,
  Todd, Isaac, Teo, Kang, Tarte, Barber, Evetts, and Blamire}}]{Mathur_DWR}
\bibinfo{author}{\bibfnamefont{N.}~\bibnamefont{Mathur}},
  \bibinfo{author}{\bibfnamefont{P.}~\bibnamefont{Littlewood}},
  \bibinfo{author}{\bibfnamefont{N.}~\bibnamefont{Todd}},
  \bibinfo{author}{\bibfnamefont{S.}~\bibnamefont{Isaac}},
  \bibinfo{author}{\bibfnamefont{B.}~\bibnamefont{Teo}},
  \bibinfo{author}{\bibfnamefont{D.}~\bibnamefont{Kang}},
  \bibinfo{author}{\bibfnamefont{E.}~\bibnamefont{Tarte}},
  \bibinfo{author}{\bibfnamefont{Z.}~\bibnamefont{Barber}},
  \bibinfo{author}{\bibfnamefont{J.}~\bibnamefont{Evetts}}, \bibnamefont{and}
  \bibinfo{author}{\bibfnamefont{M.}~\bibnamefont{Blamire}},
  \bibinfo{journal}{Journal of Applied Physics} \textbf{\bibinfo{volume}{86}},
  \bibinfo{pages}{6287} (\bibinfo{year}{1999}).

\bibitem[{\citenamefont{Suzuki et~al.}(2000)\citenamefont{Suzuki, Wu, Yu,
  Ruediger, Kent, Nath, and Eom}}]{Suzuki_DWR}
\bibinfo{author}{\bibfnamefont{Y.}~\bibnamefont{Suzuki}},
  \bibinfo{author}{\bibfnamefont{Y.}~\bibnamefont{Wu}},
  \bibinfo{author}{\bibfnamefont{J.}~\bibnamefont{Yu}},
  \bibinfo{author}{\bibfnamefont{U.}~\bibnamefont{Ruediger}},
  \bibinfo{author}{\bibfnamefont{A.}~\bibnamefont{Kent}},
  \bibinfo{author}{\bibfnamefont{T.}~\bibnamefont{Nath}}, \bibnamefont{and}
  \bibinfo{author}{\bibfnamefont{C.}~\bibnamefont{Eom}},
  \bibinfo{journal}{Journal of Applied Physics} \textbf{\bibinfo{volume}{87}},
  \bibinfo{pages}{6746} (\bibinfo{year}{2000}).

\bibitem[{\citenamefont{Yamanaka and Nagaosa}(1997)}]{Nagaosa_DWR}
\bibinfo{author}{\bibfnamefont{M.}~\bibnamefont{Yamanaka}} \bibnamefont{and}
  \bibinfo{author}{\bibfnamefont{N.}~\bibnamefont{Nagaosa}},
  \bibinfo{journal}{Physica B: Condensed Matter}
  \textbf{\bibinfo{volume}{237}}, \bibinfo{pages}{28} (\bibinfo{year}{1997}).

\bibitem[{\citenamefont{Becker et~al.}(2002)\citenamefont{Becker, Streng, Luo,
  Moshnyaga, Damaschke, Shannon, and Samwer}}]{STM_DWR}
\bibinfo{author}{\bibfnamefont{T.}~\bibnamefont{Becker}},
  \bibinfo{author}{\bibfnamefont{C.}~\bibnamefont{Streng}},
  \bibinfo{author}{\bibfnamefont{Y.}~\bibnamefont{Luo}},
  \bibinfo{author}{\bibfnamefont{V.}~\bibnamefont{Moshnyaga}},
  \bibinfo{author}{\bibfnamefont{B.}~\bibnamefont{Damaschke}},
  \bibinfo{author}{\bibfnamefont{N.}~\bibnamefont{Shannon}}, \bibnamefont{and}
  \bibinfo{author}{\bibfnamefont{K.}~\bibnamefont{Samwer}},
  \bibinfo{journal}{Phys. Rev. Lett.} \textbf{\bibinfo{volume}{89}},
  \bibinfo{pages}{237203} (\bibinfo{year}{2002}).

\bibitem[{\citenamefont{Singh-Bhalla et~al.}(2009)\citenamefont{Singh-Bhalla,
  Selcuk, Dhakal, Biswas, and Hebard}}]{FLORIDA_DWR}
\bibinfo{author}{\bibfnamefont{G.}~\bibnamefont{Singh-Bhalla}},
  \bibinfo{author}{\bibfnamefont{S.}~\bibnamefont{Selcuk}},
  \bibinfo{author}{\bibfnamefont{T.}~\bibnamefont{Dhakal}},
  \bibinfo{author}{\bibfnamefont{A.}~\bibnamefont{Biswas}}, \bibnamefont{and}
  \bibinfo{author}{\bibfnamefont{A.~F.} \bibnamefont{Hebard}},
  \bibinfo{journal}{Phys. Rev. Lett.} \textbf{\bibinfo{volume}{102}},
  \bibinfo{pages}{077205} (\bibinfo{year}{2009}).

\bibitem[{\citenamefont{Arnal et~al.}(2007)\citenamefont{Arnal, Khvalkovskii,
  Bibes, Mercey, Lecoeur, and Haghiri-Gosnet}}]{insulatingDWs}
\bibinfo{author}{\bibfnamefont{T.}~\bibnamefont{Arnal}},
  \bibinfo{author}{\bibfnamefont{A.~V.} \bibnamefont{Khvalkovskii}},
  \bibinfo{author}{\bibfnamefont{M.}~\bibnamefont{Bibes}},
  \bibinfo{author}{\bibfnamefont{B.}~\bibnamefont{Mercey}},
  \bibinfo{author}{\bibfnamefont{P.}~\bibnamefont{Lecoeur}}, \bibnamefont{and}
  \bibinfo{author}{\bibfnamefont{A.-M.} \bibnamefont{Haghiri-Gosnet}},
  \bibinfo{journal}{Phys. Rev. B} \textbf{\bibinfo{volume}{75}},
  \bibinfo{pages}{220409(R)} (\bibinfo{year}{2007}).

\bibitem[{\citenamefont{Ginzburg and Landau}(1950)}]{GinzburgLandau}
\bibinfo{author}{\bibfnamefont{V.~L.} \bibnamefont{Ginzburg}} \bibnamefont{and}
  \bibinfo{author}{\bibfnamefont{L.~D.} \bibnamefont{Landau}},
  \bibinfo{journal}{h. Eksp. Teor. Fiz.} \textbf{\bibinfo{volume}{20}},
  \bibinfo{pages}{1064} (\bibinfo{year}{1950}).

\bibitem[{\citenamefont{Streiffer et~al.}(2002)\citenamefont{Streiffer,
  Eastman, Fong, Thompson, Munkholm, Ramana~Murty, Auciello, Bai, and
  Stephenson}}]{domainsPTO}
\bibinfo{author}{\bibfnamefont{S.~K.} \bibnamefont{Streiffer}},
  \bibinfo{author}{\bibfnamefont{J.~A.} \bibnamefont{Eastman}},
  \bibinfo{author}{\bibfnamefont{D.~D.} \bibnamefont{Fong}},
  \bibinfo{author}{\bibfnamefont{C.}~\bibnamefont{Thompson}},
  \bibinfo{author}{\bibfnamefont{A.}~\bibnamefont{Munkholm}},
  \bibinfo{author}{\bibfnamefont{M.~V.} \bibnamefont{Ramana~Murty}},
  \bibinfo{author}{\bibfnamefont{O.}~\bibnamefont{Auciello}},
  \bibinfo{author}{\bibfnamefont{G.~R.} \bibnamefont{Bai}}, \bibnamefont{and}
  \bibinfo{author}{\bibfnamefont{G.~B.} \bibnamefont{Stephenson}},
  \bibinfo{journal}{Phys. Rev. Lett.} \textbf{\bibinfo{volume}{89}},
  \bibinfo{pages}{067601} (\bibinfo{year}{2002}).

\bibitem[{\citenamefont{Aguado-Puente and Junquera}(2008)}]{Junquera_closure}
\bibinfo{author}{\bibfnamefont{P.}~\bibnamefont{Aguado-Puente}}
  \bibnamefont{and} \bibinfo{author}{\bibfnamefont{J.}~\bibnamefont{Junquera}},
  \bibinfo{journal}{Phys. Rev. Lett.} \textbf{\bibinfo{volume}{100}},
  \bibinfo{pages}{177601} (\bibinfo{year}{2008}).

\bibitem[{\citenamefont{P{\v{r}}{\'\i}vratsk{\'a} and
  Janovec}(1997)}]{Privratska}
\bibinfo{author}{\bibfnamefont{J.}~\bibnamefont{P{\v{r}}{\'\i}vratsk{\'a}}}
  \bibnamefont{and} \bibinfo{author}{\bibfnamefont{V.}~\bibnamefont{Janovec}},
  \bibinfo{journal}{Ferroelectrics} \textbf{\bibinfo{volume}{204}},
  \bibinfo{pages}{321} (\bibinfo{year}{1997}).

\bibitem[{\citenamefont{Aird and Salje}(1998)}]{Arid}
\bibinfo{author}{\bibfnamefont{A.}~\bibnamefont{Aird}} \bibnamefont{and}
  \bibinfo{author}{\bibfnamefont{E.~K.~H.} \bibnamefont{Salje}},
  \bibinfo{journal}{Journal of Physics: Condensed Matter}
  \textbf{\bibinfo{volume}{10}}, \bibinfo{pages}{L377} (\bibinfo{year}{1998}).

\bibitem[{\citenamefont{Bartels et~al.}(2003)\citenamefont{Bartels, Hagen,
  Burianek, Getzlaff, Bismayer, and Wiesendanger}}]{Bartels}
\bibinfo{author}{\bibfnamefont{M.}~\bibnamefont{Bartels}},
  \bibinfo{author}{\bibfnamefont{V.}~\bibnamefont{Hagen}},
  \bibinfo{author}{\bibfnamefont{M.}~\bibnamefont{Burianek}},
  \bibinfo{author}{\bibfnamefont{M.}~\bibnamefont{Getzlaff}},
  \bibinfo{author}{\bibfnamefont{U.}~\bibnamefont{Bismayer}}, \bibnamefont{and}
  \bibinfo{author}{\bibfnamefont{R.}~\bibnamefont{Wiesendanger}},
  \bibinfo{journal}{Journal of Physics: Condensed Matter}
  \textbf{\bibinfo{volume}{15}}, \bibinfo{pages}{957} (\bibinfo{year}{2003}).

\bibitem[{\citenamefont{Fiebig et~al.}(2002)\citenamefont{Fiebig, Lottermoser,
  Fr{q\"o}hlich, Goltsev, and Pisarev}}]{Fiebig2002clampedDWs}
\bibinfo{author}{\bibfnamefont{M.}~\bibnamefont{Fiebig}},
  \bibinfo{author}{\bibfnamefont{T.}~\bibnamefont{Lottermoser}},
  \bibinfo{author}{\bibfnamefont{D.}~\bibnamefont{Fr{q\"o}hlich}},
  \bibinfo{author}{\bibfnamefont{A.}~\bibnamefont{Goltsev}}, \bibnamefont{and}
  \bibinfo{author}{\bibfnamefont{R.}~\bibnamefont{Pisarev}},
  \bibinfo{journal}{Nature} \textbf{\bibinfo{volume}{419}},
  \bibinfo{pages}{818} (\bibinfo{year}{2002}).

\bibitem[{\citenamefont{Lottermoser and
  Fiebig}(2004)}]{Lottermoser2004DWsHoMnO3}
\bibinfo{author}{\bibfnamefont{T.}~\bibnamefont{Lottermoser}} \bibnamefont{and}
  \bibinfo{author}{\bibfnamefont{M.}~\bibnamefont{Fiebig}},
  \bibinfo{journal}{Phys. Rev. B} \textbf{\bibinfo{volume}{70}},
  \bibinfo{pages}{220407(R)} (\bibinfo{year}{2004}).

\bibitem[{\citenamefont{Kagawa et~al.}(2009)\citenamefont{Kagawa, Mochizuki,
  Onose, Murakawa, Kaneko, Furukawa, and Tokura}}]{Tokura2002dynamicsDW}
\bibinfo{author}{\bibfnamefont{F.}~\bibnamefont{Kagawa}},
  \bibinfo{author}{\bibfnamefont{M.}~\bibnamefont{Mochizuki}},
  \bibinfo{author}{\bibfnamefont{Y.}~\bibnamefont{Onose}},
  \bibinfo{author}{\bibfnamefont{H.}~\bibnamefont{Murakawa}},
  \bibinfo{author}{\bibfnamefont{Y.}~\bibnamefont{Kaneko}},
  \bibinfo{author}{\bibfnamefont{N.}~\bibnamefont{Furukawa}}, \bibnamefont{and}
  \bibinfo{author}{\bibfnamefont{Y.}~\bibnamefont{Tokura}},
  \bibinfo{journal}{Phys. Rev. Lett.} \textbf{\bibinfo{volume}{102}},
  \bibinfo{pages}{057604} (\bibinfo{year}{2009}).

\bibitem[{\citenamefont{Seidel et~al.}(2009)\citenamefont{Seidel, Martin, He,
  Zhan, Chu, Rother, Hawkridge, Maksymovych, Yu, Gajek et~al.}}]{Seidel}
\bibinfo{author}{\bibfnamefont{J.}~\bibnamefont{Seidel}},
  \bibinfo{author}{\bibfnamefont{L.}~\bibnamefont{Martin}},
  \bibinfo{author}{\bibfnamefont{Q.}~\bibnamefont{He}},
  \bibinfo{author}{\bibfnamefont{Q.}~\bibnamefont{Zhan}},
  \bibinfo{author}{\bibfnamefont{Y.}~\bibnamefont{Chu}},
  \bibinfo{author}{\bibfnamefont{A.}~\bibnamefont{Rother}},
  \bibinfo{author}{\bibfnamefont{M.}~\bibnamefont{Hawkridge}},
  \bibinfo{author}{\bibfnamefont{P.}~\bibnamefont{Maksymovych}},
  \bibinfo{author}{\bibfnamefont{P.}~\bibnamefont{Yu}},
  \bibinfo{author}{\bibfnamefont{M.}~\bibnamefont{Gajek}},
  \bibnamefont{et~al.}, \bibinfo{journal}{Nature Materials}
  \textbf{\bibinfo{volume}{8}}, \bibinfo{pages}{229} (\bibinfo{year}{2009}).

\bibitem[{\citenamefont{Choi et~al.}(2010)\citenamefont{Choi, Horibe, Yi, Choi,
  Wu, and Cheong}}]{Choi}
\bibinfo{author}{\bibfnamefont{T.}~\bibnamefont{Choi}},
  \bibinfo{author}{\bibfnamefont{Y.}~\bibnamefont{Horibe}},
  \bibinfo{author}{\bibfnamefont{H.}~\bibnamefont{Yi}},
  \bibinfo{author}{\bibfnamefont{Y.}~\bibnamefont{Choi}},
  \bibinfo{author}{\bibfnamefont{W.}~\bibnamefont{Wu}}, \bibnamefont{and}
  \bibinfo{author}{\bibfnamefont{S.}~\bibnamefont{Cheong}},
  \bibinfo{journal}{Nature Materials} \textbf{\bibinfo{volume}{9}},
  \bibinfo{pages}{253} (\bibinfo{year}{2010}).

\bibitem[{\citenamefont{Hotta et~al.}(2003)\citenamefont{Hotta, Moraghebi,
  Feiguin, Moreo, Yunoki, and Dagotto}}]{hotta_2003a}
\bibinfo{author}{\bibfnamefont{T.}~\bibnamefont{Hotta}},
  \bibinfo{author}{\bibfnamefont{M.}~\bibnamefont{Moraghebi}},
  \bibinfo{author}{\bibfnamefont{A.}~\bibnamefont{Feiguin}},
  \bibinfo{author}{\bibfnamefont{A.}~\bibnamefont{Moreo}},
  \bibinfo{author}{\bibfnamefont{S.}~\bibnamefont{Yunoki}}, \bibnamefont{and}
  \bibinfo{author}{\bibfnamefont{E.}~\bibnamefont{Dagotto}},
  \bibinfo{journal}{Phys.\ Rev.\ Lett.} \textbf{\bibinfo{volume}{90}},
  \bibinfo{pages}{247203} (\bibinfo{year}{2003}).

\bibitem[{\citenamefont{Rodr\'\i{}guez-Carvajal
  et~al.}(1998)\citenamefont{Rodr\'\i{}guez-Carvajal, Hennion, Moussa, Moudden,
  Pinsard, and Revcolevschi}}]{rodriguezcarvajalLaMnO3}
\bibinfo{author}{\bibfnamefont{J.}~\bibnamefont{Rodr\'\i{}guez-Carvajal}},
  \bibinfo{author}{\bibfnamefont{M.}~\bibnamefont{Hennion}},
  \bibinfo{author}{\bibfnamefont{F.}~\bibnamefont{Moussa}},
  \bibinfo{author}{\bibfnamefont{A.~H.} \bibnamefont{Moudden}},
  \bibinfo{author}{\bibfnamefont{L.}~\bibnamefont{Pinsard}}, \bibnamefont{and}
  \bibinfo{author}{\bibfnamefont{A.}~\bibnamefont{Revcolevschi}},
  \bibinfo{journal}{Phys. Rev. B} \textbf{\bibinfo{volume}{57}},
  \bibinfo{pages}{R3189} (\bibinfo{year}{1998}).

\bibitem[{\citenamefont{Dagotto}(2002)}]{Dagotto_book}
\bibinfo{author}{\bibfnamefont{E.}~\bibnamefont{Dagotto}},
  \emph{\bibinfo{title}{Nanoscale Phase Separation and Colossal
  Magnetoresistance}} (\bibinfo{publisher}{Springer-Verlag},
  \bibinfo{address}{Berlin}, \bibinfo{year}{2002}).

\bibitem[{\citenamefont{Slater and Koster}(1954)}]{SlaterKoster}
\bibinfo{author}{\bibfnamefont{J.~C.} \bibnamefont{Slater}} \bibnamefont{and}
  \bibinfo{author}{\bibfnamefont{G.~F.} \bibnamefont{Koster}},
  \bibinfo{journal}{Phys. Rev.} \textbf{\bibinfo{volume}{94}},
  \bibinfo{pages}{1498} (\bibinfo{year}{1954}).

\bibitem[{\citenamefont{Salafranca et~al.}(2009)\citenamefont{Salafranca,
  Alvarez, and Dagotto}}]{Fermiarcs}
\bibinfo{author}{\bibfnamefont{J.}~\bibnamefont{Salafranca}},
  \bibinfo{author}{\bibfnamefont{G.}~\bibnamefont{Alvarez}}, \bibnamefont{and}
  \bibinfo{author}{\bibfnamefont{E.}~\bibnamefont{Dagotto}},
  \bibinfo{journal}{Phys. Rev. B} \textbf{\bibinfo{volume}{80}},
  \bibinfo{pages}{155133} (\bibinfo{year}{2009}).

\bibitem[{\citenamefont{Hotta et~al.}(1999)\citenamefont{Hotta, Yunoki, Mayr,
  and Dagotto}}]{Hotta_hartree}
\bibinfo{author}{\bibfnamefont{T.}~\bibnamefont{Hotta}},
  \bibinfo{author}{\bibfnamefont{S.}~\bibnamefont{Yunoki}},
  \bibinfo{author}{\bibfnamefont{M.}~\bibnamefont{Mayr}}, \bibnamefont{and}
  \bibinfo{author}{\bibfnamefont{E.}~\bibnamefont{Dagotto}},
  \bibinfo{journal}{Phys. Rev. B} \textbf{\bibinfo{volume}{60}},
  \bibinfo{pages}{R15009} (\bibinfo{year}{1999}).

\bibitem[{\citenamefont{Wollan and Koehler}(1955)}]{Wollan}
\bibinfo{author}{\bibfnamefont{E.}~\bibnamefont{Wollan}} \bibnamefont{and}
  \bibinfo{author}{\bibfnamefont{W.}~\bibnamefont{Koehler}},
  \bibinfo{journal}{Phys.\ Rev.} \textbf{\bibinfo{volume}{100}},
  \bibinfo{pages}{545} (\bibinfo{year}{1955}).

\bibitem[{\citenamefont{Anderson et~al.}(1999)\citenamefont{Anderson, Bai,
  Bischof, Blackford, Demmel, Dongarra, Du~Croz, Greenbaum, Hammarling,
  McKenney et~al.}}]{lapack}
\bibinfo{author}{\bibfnamefont{E.}~\bibnamefont{Anderson}},
  \bibinfo{author}{\bibfnamefont{Z.}~\bibnamefont{Bai}},
  \bibinfo{author}{\bibfnamefont{C.}~\bibnamefont{Bischof}},
  \bibinfo{author}{\bibfnamefont{S.}~\bibnamefont{Blackford}},
  \bibinfo{author}{\bibfnamefont{J.}~\bibnamefont{Demmel}},
  \bibinfo{author}{\bibfnamefont{J.}~\bibnamefont{Dongarra}},
  \bibinfo{author}{\bibfnamefont{J.}~\bibnamefont{Du~Croz}},
  \bibinfo{author}{\bibfnamefont{A.}~\bibnamefont{Greenbaum}},
  \bibinfo{author}{\bibfnamefont{S.}~\bibnamefont{Hammarling}},
  \bibinfo{author}{\bibfnamefont{A.}~\bibnamefont{McKenney}},
  \bibnamefont{et~al.}, \emph{\bibinfo{title}{{LAPACK} Users' Guide}}
  (\bibinfo{publisher}{Society for Industrial and Applied Mathematics},
  \bibinfo{address}{Philadelphia, PA}, \bibinfo{year}{1999}),
  \bibinfo{edition}{3rd} ed.

\bibitem[{\citenamefont{Salafranca and Verg\'es}(2006)}]{SalafrancaMonte}
\bibinfo{author}{\bibfnamefont{J.}~\bibnamefont{Salafranca}} \bibnamefont{and}
  \bibinfo{author}{\bibfnamefont{J.~A.} \bibnamefont{Verg\'es}},
  \bibinfo{journal}{Phys. Rev. B} \textbf{\bibinfo{volume}{74}},
  \bibinfo{pages}{184428} (\bibinfo{year}{2006}).

\bibitem[{\citenamefont{{Press} et~al.}(1992)\citenamefont{{Press},
  {Teukolsky}, {Vetterling}, and {Flannery}}}]{Recipes}
\bibinfo{author}{\bibfnamefont{W.~H.} \bibnamefont{{Press}}},
  \bibinfo{author}{\bibfnamefont{S.~A.} \bibnamefont{{Teukolsky}}},
  \bibinfo{author}{\bibfnamefont{W.~T.} \bibnamefont{{Vetterling}}},
  \bibnamefont{and} \bibinfo{author}{\bibfnamefont{B.~P.}
  \bibnamefont{{Flannery}}}, \emph{\bibinfo{title}{{Numerical recipes in
  FORTRAN. The art of scientific computing}}} (\bibinfo{year}{1992}).

\bibitem[{\citenamefont{\ifmmode~\mbox{\c{S}}\else \c{S}\fi{}en
  et~al.}(2004)\citenamefont{\ifmmode~\mbox{\c{S}}\else \c{S}\fi{}en, Alvarez,
  and Dagotto}}]{Sen_04}
\bibinfo{author}{\bibfnamefont{C.}~\bibnamefont{\ifmmode~\mbox{\c{S}}\else
  \c{S}\fi{}en}}, \bibinfo{author}{\bibfnamefont{G.}~\bibnamefont{Alvarez}},
  \bibnamefont{and} \bibinfo{author}{\bibfnamefont{E.}~\bibnamefont{Dagotto}},
  \bibinfo{journal}{Phys. Rev. B} \textbf{\bibinfo{volume}{70}},
  \bibinfo{pages}{064428} (\bibinfo{year}{2004}).

\bibitem[{\citenamefont{Salafranca and Brey}(2006)}]{incommensurate}
\bibinfo{author}{\bibfnamefont{J.}~\bibnamefont{Salafranca}} \bibnamefont{and}
  \bibinfo{author}{\bibfnamefont{L.}~\bibnamefont{Brey}},
  \bibinfo{journal}{Phys. Rev. B} \textbf{\bibinfo{volume}{73}},
  \bibinfo{pages}{024422} (\bibinfo{year}{2006}).

\bibitem[{\citenamefont{Verg{\'e}s}(1999)}]{VergesCPC}
\bibinfo{author}{\bibfnamefont{J.}~\bibnamefont{Verg{\'e}s}},
  \bibinfo{journal}{Comput.\ Phys.\ Commun.} \textbf{\bibinfo{volume}{118}},
  \bibinfo{pages}{71} (\bibinfo{year}{1999}).

\bibitem[{\citenamefont{Popovic and Satpathy}(2000)}]{Satpathy}
\bibinfo{author}{\bibfnamefont{Z.}~\bibnamefont{Popovic}} \bibnamefont{and}
  \bibinfo{author}{\bibfnamefont{S.}~\bibnamefont{Satpathy}},
  \bibinfo{journal}{Phys. Rev. Lett.} \textbf{\bibinfo{volume}{84}},
  \bibinfo{pages}{1603} (\bibinfo{year}{2000}).

\bibitem[{\citenamefont{Murakami et~al.}(2006)\citenamefont{Murakami, Yamauchi,
  Nakamura, Moritomo, Tanaka, and Kawai}}]{gapLMO1}
\bibinfo{author}{\bibfnamefont{K.}~\bibnamefont{Murakami}},
  \bibinfo{author}{\bibfnamefont{T.}~\bibnamefont{Yamauchi}},
  \bibinfo{author}{\bibfnamefont{A.}~\bibnamefont{Nakamura}},
  \bibinfo{author}{\bibfnamefont{Y.}~\bibnamefont{Moritomo}},
  \bibinfo{author}{\bibfnamefont{H.}~\bibnamefont{Tanaka}}, \bibnamefont{and}
  \bibinfo{author}{\bibfnamefont{T.}~\bibnamefont{Kawai}},
  \bibinfo{journal}{Phys. Rev. B} \textbf{\bibinfo{volume}{73}},
  \bibinfo{pages}{180403(R)} (\bibinfo{year}{2006}).

\bibitem[{\citenamefont{Inami et~al.}(2003)\citenamefont{Inami, Fukuda, Mizuki,
  Ishihara, Kondo, Nakao, Matsumura, Hirota, Murakami, Maekawa
  et~al.}}]{gapLMO2}
\bibinfo{author}{\bibfnamefont{T.}~\bibnamefont{Inami}},
  \bibinfo{author}{\bibfnamefont{T.}~\bibnamefont{Fukuda}},
  \bibinfo{author}{\bibfnamefont{J.}~\bibnamefont{Mizuki}},
  \bibinfo{author}{\bibfnamefont{S.}~\bibnamefont{Ishihara}},
  \bibinfo{author}{\bibfnamefont{H.}~\bibnamefont{Kondo}},
  \bibinfo{author}{\bibfnamefont{H.}~\bibnamefont{Nakao}},
  \bibinfo{author}{\bibfnamefont{T.}~\bibnamefont{Matsumura}},
  \bibinfo{author}{\bibfnamefont{K.}~\bibnamefont{Hirota}},
  \bibinfo{author}{\bibfnamefont{Y.}~\bibnamefont{Murakami}},
  \bibinfo{author}{\bibfnamefont{S.}~\bibnamefont{Maekawa}},
  \bibnamefont{et~al.}, \bibinfo{journal}{Phys. Rev. B}
  \textbf{\bibinfo{volume}{67}}, \bibinfo{pages}{045108}
  (\bibinfo{year}{2003}).

\end{thebibliography}

\end{document}